\documentstyle[natbib2, epsfig,natbibmnfix,epstopdf]{mn2e}
\def\xmm{{\sl XMM-Newton}}
\def\rms{$\sigma_{\rm rms}$}
\def\nxs{$\sigma^2_{\rm NXS}$}

\def\mkn{{Mkn~766}}
\title[Fourier-resolved spectra of Mkn 766]{Fourier-resolved energy spectra of the Narrow Line Seyfert 1 Mkn 766}  
\author[P. Ar\'evalo et al.]{P. Ar\'evalo$^{1}$\thanks{E-mail: patricia@astro.soton.ac.uk}, I. M. McHardy$^{1}$, A. Markowitz$^{2}$, I. E. Papadakis$^{3}$, T. J. Turner$^{4,5}$,\and  L. Miller$^{6}$, and J. Reeves$^{7}$ \\ 
$^1$School of Physics and Astronomy, University of Southampton, Southampton SO17 1BJ, UK\\
$^2$Center for Astrophysics and Space Sciences, University of California,
San Diego, Mail Code 0424,  La Jolla, CA 92093-0424, USA\\
$^3$Physics Department, University of Crete, P.O. box 2208, 71003 Heraklion, Greece\\
$^4$Department of Physics, University of Maryland Baltimore County, 1000 Hilltop Circle, Baltimore, MD 21250, USA\\ 
$^5$Code 662,  Exploration of the Universe Division, NASA/GSFC, Greenbelt, MD 20771, USA\\
$^6$Dept. of Physics, University of Oxford, Denys Wilkinson Building, Keble Road, Oxford OX1 3RH, UK\\
$^7$ Astrophysics Group, School of Physics and Geographical Sciences, Keele University, Keele, Staffordshire ST5 5BG, UK \\ 
}

\begin{document}
\date{Received /Accepted}
\pagerange{\pageref{firstpage}--\pageref{lastpage}} \pubyear{2006}

\maketitle
\label{firstpage}

\begin{abstract}

We compute Fourier-resolved X-ray spectra of the Seyfert 1 Markarian
766 to study the shape of the variable components contributing to the
0.3--10 keV energy spectrum and their time-scale dependence. The
fractional variability spectra peak at 1--3 keV, as in other Seyfert 1
galaxies, consistent with either a constant contribution from a soft
excess component below 1 keV and Compton reflection component above 2
keV, or variable warm absorption enhancing the variability in the 1--3
keV range.  The rms spectra, which shows the shape of the variable
components only, is well described by a single power law with an
absorption feature around 0.7 keV, which gives it an apparent soft
excess. This spectral shape can be produced by a power law varying in
normalisation, affected by an approximately constant (within each
orbit) warm absorber, with parameters similar to those found by Turner
et al. for the warm-absorber layer covering all spectral components in
their scattering scenario ($N_H\sim 3 \times 10^{21} {\rm cm} ^{-2},
\log(\xi)\sim 1$). The total soft excess in the average spectrum can
therefore be produced by a combination of constant warm absorption on
the power law plus an additional less variable component. On shorter
time-scales, the rms spectrum hardens and this evolution is well
described by a change in power law slope, while the absorption
parameters remain the same. The frequency dependence of the rms
spectra can be interpreted as variability arising from propagating
fluctuations through an extended emitting region, whose emitted
spectrum is a power law that hardens towards the centre. This scenario
reduces the short time-scale variability of lower energy bands making
the variable spectrum harder on shorter time-scales and at the same
time explains the hard lags found in these data by Markowitz et al.
\end{abstract}


\section{Introduction}

The X-ray emission of active galactic nuclei (AGN) is thought to be
produced in the innermost regions of these systems, close to the
central super massive black hole.  The X-ray spectrum of most Narrow
Line Seyfert 1 galaxies above 1 keV can be broadly interpreted as a
power law component, probably arising from inverse Compton scattering
of thermal disc photons through an optically thin corona, plus a
Compton reflection component peaking towards $\sim$ 40 keV. In Narrow
Line Seyfert 1s (NLS1s), a soft excess appears below $\sim 1$ keV,
above an extrapolation of the power law component. This soft excess
has been interpreted as either additional emission, e.g. a thermal
black body or relativistically blurred ionized reflection
\citep{fabian02,crummy} or as warm absorption on the power law
component, that produces an apparent excess at energies below the
absorption edges \citep{gierlinski}. Spectral fitting on its own
cannot distinguish between these possibilities and indeed, gives
little information on the geometry of the emitting regions. Combining
spectral fitting with a study of the variability properties of the
different energy bands gives information on the connection between the
spectral components and their possible emission mechanisms.

The fractional variability of the X-ray light curves is energy
dependent. In the case of Narrow Line Seyfert 1s, such as \mkn , the
variability appears strongest in the middle of the \xmm\ band, around
1--2 keV decreasing noticeably towards lower and higher energies. This
behaviour suggests that the variability is produced mainly by the
power law component while the soft excess and Compton reflection
components remain constant \citep[e.g][]{fabian02}. In this scenario,
the constant components add to the flux of the variable component but
not to the variability, thus diluting the fractional variability
in the energy bands where they dominate the spectrum. This
two-component (constant plus variable) model can broadly reproduce the
flux-dependent spectral changes observed in several AGN \citep{taylor,
  vaughan04} . In particular, \citet{miller} have used the available
$\sim 600$ ks of \xmm\ data of \mkn\ to show that the spectral
variability of this AGN on time-scales longer than 20 ks, can be
largely explained through two distinct spectral components that vary
in relative normalisation.

In its simplest form, however, the two-component model cannot
reproduce the \emph{time-scale} dependence of the spectral
variability, i.e. if only one of the spectral components varies in
normalisation, the spectral changes would depend only on the relative
fluxes regardless of the time-scale of their fluctuations.  The fact
that the energy bands behave differently on different time-scales can
be readily seen through the power density spectrum (PDS), which
measures the variability power as a function of Fourier frequency. In
most cases, the PDS of AGN are well described by a broken power law
model, with a slope of $\sim -1$ below the break and $\sim -2$ or
steeper, at higher frequencies.  Energy dependence of the PDS shape is
normally observable around the break frequency and implies that higher
energy bands are more rapidly varying. This additional variability
power can appear as a flatter high-frequency PDS slope for higher
energies, as in NGC~4051 \citep{McHardy4051}, as a shift of the break
to higher frequencies, as observed for the first time in \mkn\ by
\citet{markowitz} or as a high-frequency Lorentzian component that
appears stronger in the high-energy PDS of Ark~564 \citep{mchardy07}.

This simple two-component model can explain the energy dependence of
the PDS normalisation but not energy-dependent PDS shapes. The fact
that many AGN do display more short time-scale variability at higher
energies does not rule out two-component models, but it implies that
the variability is more complex than simple changes in normalisation
of one of the components. One possibility is that the variable
component changes its spectral shape when viewed on different
time-scales. This can happen, for example, if softer energies are
emitted by a more extended region, then their short time-scale
variability can be suppressed and, therefore, the high-frequency
spectrum appears harder. Alternatively, both spectral components might
be variable but with different timing-properties, each one defining the
behaviour of the energy bands where they dominate the spectrum.

\citet{markowitz} studied in detail the energy dependence of the PDS
of \mkn\ using \xmm\ data. They find that the 1.1--12 keV energy band
has significantly more variability power than lower energy bands, on
time-scales around the break in the PDS. This energy-dependent shape
can be either interpreted as a shift of the break frequency to higher
values for higher energies or as an additional variability component
with a hard energy spectrum and a band-limited PDS. In this paper, we
analyze the same \xmm\ data set to determine the dependence of the
amplitude of the variability on energy and time-scale by calculating
the Fourier-resolved spectra \citep{Revnivtsev}. In short, this
technique produces the absolute root-mean-square (rms) amplitude of
variability and also the fractional rms variability (rms divided by
count rate) as a function of energy and time-scale. If the variability
is produced by a spectral component varying in normalisation,
the Fourier-resolved spectrum will have the same shape as the variable
spectral component. If there are additional constant spectral
components, these will dilute the \emph{fractional} variability in the
energy bands where this constant component dominates. The dilution of
the fractional variability is evidenced in the \emph{normalised}
variance spectra.

The paper is organized as follows: we describe the
data reduction in Sec. \ref{data}, and the Fourier-resolved spectrum
technique in Sec. \ref{technique}. The resulting normalised excess
variance (\nxs ) spectra, discussed in Sec. \ref{nxs} show the
characteristic shape found in other NLS1s where the \nxs\ peaks around
1--2 keV, at low frequencies. At high frequencies however, the \nxs\
spectra becomes harder. This frequency dependence is studied in more
detail in Sec. \ref{rms} where we examine the spectral shape of the
variable components only, through the unnormalised \rms\ spectra. We
discuss the possible contribution of an additional spectral component
to the \rms\ spectra in Sec. \ref{constant}. In Sec. \ref{discussion}
we discuss the origin of the variability and frequency dependence of
the \rms\ spectra in different possible scenarios.

\section{The data}
\label{data}
\mkn\ was recently observed by \xmm\ for $\sim 500$ ks from 2005-05-23
to 2005-05-31 during revolutions 999--1004 (observation ID in the
range 030403[1-7]01). Here we combine this data set with
earlier observations made on 2000-05-20 (obs ID 0096020101) and
2001-05-20--21 (obs ID 0109141301) during orbits 82 and 265,
respectively. Spectral fitting and spectral variability analyses of these data have been published by \citet{miller06, miller,turner,turner06}.

\subsection{Observations}

We used data from the EPIC PN detector \citep{struder}. In all
observations, the PN camera was operated in Small Window mode, using
medium filter. The data were processed using XMM-SAS v6.5.0.  For each
orbit, we extracted photons from a circular region of $5\arcmin$ in
radius, centered on the source, and chose a background region of equal
area on the same chip. Source and background events were selected by
quality flag=0 and patterns=0--4, i.e. we kept only single and double
events. The background level was generally low and stable, except at
the beginning and end of each orbit and for $\sim 1$ ks in the middle
of orbit 82 and $\sim 4 $ ks in the middle of orbit 1002.

\subsection{Light curves}
\label{seclc}
We made light curves in 19 energy bands: 6 bands of width 0.1 keV,
between 0.2 and 0.8 keV, then the following bands: 0.8--1.0, 1.0--1.3,
1.3--1.6, 1.6--2 keV, 4 bands of width 0.5 keV between 2 and 4 keV, 4
bands of width 1 keV between 4 and 8 keV, and a final band 8--10
keV. The width of the energy bands, which will define the energy
resolution of the FR spectra, was chosen to keep enough counts per bin
to produce reliable power spectra. The light curves were binned in 100
s bins and the background light curve were subtracted. After removing
times of high background activity 
the final light curves had lengths of 36, 105 ,77, 96, 93, 93 and 85
ks, for orbits 82, 265, 999, 1000, 1001, 1002 and 1003
respectively. The 4 ks gap in the light curves of orbit 1002 and a 1
ks gap in orbit 82 were filled in by interpolating and adding Poisson
noise deviates at the appropriate levels. We discarded the data from
orbit 1004 because of its short duration as it is not long enough to
cover the low frequency range used in this analysis.

Response matrices and ancillary response files were created using the
XMM-SAS tasks rmfgen and arfgen, separately for orbits 82, 265 and
999-1003. The response matrices were later re-binned in energy using
the task rbnrmf to match the energy bands used for the light curves.

\section{Fourier-resolved spectrum technique} 
\label{technique}
The Fourier-resolved spectrum (FR spectrum) technique, developed by
\citet{Revnivtsev}, calculates the root-mean-square spread, $\sigma_{\rm rms}$, of different
energy bands as a function of variability time-scale.  This
technique was applied to AGN data for the first time by
\citet{Papadakis_frspec}, who studied the spectral variability of the
NLS1s MCG--6-30-15 and later by \citet{Papadakis_frspec2}, who
studied \mkn\ (among other AGN), using the orbit 265 data only and
restricted the analysis to the high energy band (3--10 keV). In all
cases, the FR spectra resulted to be noticeably frequency-dependent,
with higher Fourier frequencies displaying harder spectra.

The FR spectrum is calculated through the power density spectrum
(PDS), which measures the variability power in different Fourier
frequencies. For a discrete time series $x(t_j)$, the normalised PDS
is obtained through the discrete Fourier transform (DFT, Press et
al. 1996) as:
\begin{equation} 
P(f_i)=\frac{2\Delta T}{N \bar x ^2} |{\rm DFT}(f_i)|^2=\frac{2\Delta
T}{N \bar x ^2}\left|\sum^{N}_{j=1} x_j e^{2\pi i f_i t_j}\right|^2
\end{equation}
where N is the number of points in the light curve, $\Delta T$ is the
time bin size, $\bar{x}$ is the average count rate, and the Fourier
frequencies are $f_i=i/T$, where $T$ is the length of the light curve.

The PDS, normalised by count rate, is given in terms of $\frac{\sigma^2}{\bar x^2} \frac{ 1}{\rm Hz}$, which
has units of 1/Hz. When the PDS is integrated over a frequency range
$\Delta f$, it produces the contribution of this frequency range to
the total normalised variance $\sigma^2_{\rm N}$. In our case, the integration is substituted by a sum of the discrete $P(f_i)$ for the Fourier frequencies (separated by $\delta f = 1/T$), within a frequency range $\Delta f=b/T-a/T$ for integers $a$ and $b$:
\begin{equation}
\sigma^2_{{\rm N,} \Delta f}=\sum^{b}_{i=a}P(f_i)\delta f.
\end{equation}

This variance will normally be the sum of the intrinsic variability
power of the source plus variability power caused by the Poisson noise
associated with the counting statistics. For a continuously sampled
light curve, this Poisson noise contributes a constant component to
the PDS at a level of ${\rm PN_{lev}}=2(\bar{x}+B)/\bar{x}^2$, where
$B$ is the average count rate of the background. The normalised excess
variance, $\sigma^2_{{\rm NXS,} \Delta f}$, produced intrinsically by the source,
in a given frequency range, $\Delta f$, can be computed by subtracting
${\rm PN_{lev}}$ from the $P(f_i)$ before summing it over $\Delta f$,
\begin{equation}
\sigma^2_{{\rm NXS,} \Delta f}=\sum^{b}_{i=a}(P(f_i)-{\rm PN_{lev}})\delta f.
\end{equation}

The normalised excess variance as a function of energy, $\sigma^2_{{\rm NXS,} \Delta f}(E)$, is obtained simply by calculating
$\sigma^2_{{\rm NXS,}\Delta f}$ for a series of adjacent energy bands.

The Poisson-noise subtracted rms variability, $\sigma_{{\rm rms,}\Delta f}$, is not normalised
by count rate and is related to $\sigma^2_{{\rm NXS,}\Delta f}$ by
\begin{equation}
\label{formulasigma}
\sigma_{{\rm rms,}\Delta f}=\bar x \sqrt{\sigma^2_{{\rm NXS,}\Delta f}}
\end{equation}
Calculating $\sigma_{{\rm rms,}\Delta f}$ for a set of energy bands
produces the $\sigma_{{\rm rms,}\Delta f}(E)$ spectrum of the
corresponding frequency range. From here on we will drop the explicit
dependence on frequency range (i.e. $\Delta f$) and energy $E$, of the
Fourier-resolved spectrum $\sigma_{{\rm rms,}\Delta f}(E)$ and of the
normalised excess variance spectrum $\sigma^2_{{\rm NXS,} \Delta f}(E)$,
writing simply \rms\ and \nxs . For a given frequency range, the
Fourier-resolved spectrum, \rms , shows the root-mean-square amplitude
of variability (in counts/s) as a function of energy, rather than the
total count rate as a function of energy which composes the
(time-averaged) energy spectrum.

Notice that if the amplitude of variability of each energy band is
proportional to its flux (i.e. the energy spectrum, $F(E)$, only
varies in normalisation but not in shape), then the $\sigma^2_{\rm
  NXS}$ spectrum will be flat, otherwise, if an energy range contains
a constant component, this will appear as a dip in the $\sigma^2_{\rm
  NXS}$ spectrum. On the other hand, the \rms\ spectrum shows the
unnormalised rms variability in each energy band, which is
proportional to the count rate \emph {and} fractional variability of
the components of the energy spectrum and so it is unaffected by any
constant spectral components. For a detailed discussion on the interpretation of the \rms\ spectrum, see \cite{Papadakis_frspec2}.

\subsection{Calculating the Fourier-resolved spectra of \mkn }  

\citet{Vaughan766} found a break in the broad-band PDS of \mkn\ at
$3\times 10^{-4}$ Hz, where the PDS slope changes from $\sim -1$ to
$\sim -2$. Markowitz et al. (2007) show that the PDS is energy
dependent, with more high frequency variability power at high
energies. This extra power represents either an increase in the break
frequency with energy or an additional band-limited variability
component, peaking at $4.6\times 10^{-4}$ Hz, and with a hard energy
spectrum \citep{markowitz}. For the Fourier-resolved spectra, we chose
three frequency bands to cover different parts of the PDS:
$10^{-4.6}-10^{-4}$ Hz, $10^{-4}-10^{-3.5}$ Hz and $
10^{-3.5}-10^{-3.0}$ Hz, which will be referred to as low-, medium-
and high-frequency, or LF, MF and HF, ranges. The LF spectrum probes
time-scales well below the break frequency, the MF probes time-scales
up to this break frequency in the low energy bands and HF covers the
frequency range where either the break frequency shifts with energy or
where the additional variability component is located. Our aim is to
determine which spectral component produces
this additional high-frequency power in the high-energy bands.

For each orbit and energy band defined in Sec. \ref{seclc}, we
calculated \nxs\ for theses three frequency ranges. Then, for each
frequency range, we collected the different energy-band variances to
produce $\sigma^2_{{\rm NXS}} (E)$ of each orbit. This dimensionless
spectrum shows the variance normalised by the flux of each energy
band. It is therefore sensitive to both constant and variable
components of the spectrum and it is broadly independent of the energy
response of the detector.

The $\sigma_{{\rm rms}}(E)$ spectra were calculated from the
$\sigma^2_{{\rm NXS}} (E)$ spectra following Eq. \ref{formulasigma},
where the $\bar x$ used were the average count rates of the
corresponding orbit and energy band. The raw shape of the
\rms\ spectra depends on the response of the detector and needs to be
fitted using the same response matrices as the total energy spectrum.

We averaged the \nxs\ spectra of orbits 999--1003 and 1000--1003, and
then multiplied by the time-average energy spectra of the
corresponding group of observations (i.e. 999--1003 or 1000--1003), to
obtain high quality \rms\ spectra. As will be shown below, the
variability behaviour of orbit 999 data is qualitatively different to
the rest of the observations, so we will also analyse it
separately. We did not combine orbit 999--1003 data with earlier
observations to avoid problems with the change in energy response of
the detector over the four years between the observations. The data
from orbit 265 produces high quality \rms\ spectra on their own, owing
to the high count rate of the source during that observation. We will
fit the \rms\ spectra of orbit 265 independently, to compare with the
behaviour of the combined new data. Orbit 82 produced lower quality
\rms\ spectra due to the relatively short exposure, so we will only
use this data for the comparison with single-orbit fits.

\subsection{Error formulae for the frequency-resolved $\sigma^2_{\rm NXS}$ and \rms\ spectra}

X-ray light curves from AGN have the variability properties of a
stochastic red-noise process. This type of process produces PDS
estimates distributed with a variance equal to the mean of  the
underlying PDS function. The error on the variance measured in a
frequency range is normally estimated from the scatter of PDS
estimates in the corresponding frequency bin (e.g. Papadakis \&
Lawrence 1993). If different energy bands are well correlated, their
binned PDS estimates will also be well correlated and so the errors
calculated from the spread of PDS estimates will greatly overestimate
the \emph{relative} errors between different energy bands, when they
are observed simultaneously. This means that the noise nature of
the light curves produces a large uncertainty in the \emph{amplitude}
of the FR spectra, but not in its \emph{shape}.

For perfectly correlated energy bands, the only source of uncertainty
in their relative PDS estimates is the uncorrelated Poisson noise in
the light curves. \citet{Vaughan_err} calculated the error expected
for $\sigma^2_{\rm NXS}$ of a red-noise light curve affected by
Poisson noise (eq. 11 in that paper), which is appropriate for the
relative scatter in variance between different energy bands. This is
only a lower limit for the errors however, as the variability of
different energy bands is not completely coherent.  We adapted the
error formulae of \citet{Vaughan_err} to produce errors for given
frequency ranges, rather than for the entire power spectrum, by
replacing the (unnormalised) variance due to Poisson noise,
$\sigma^2_{\rm err}$, by the fraction of this variance contributed by
a specific frequency range $\Delta f$, i.e. $\sigma^2_{{\rm}err,
\Delta f}={\rm PN_{lev}}\times\Delta f\times\bar x^2$.  The error
formula for \rms\ uses the error on the $\sigma^2_{\rm NXS}$ and the
conversion formula suggested by e.g. \citet{poutanen}. The resulting
error formulae are:

\begin{equation}
{\rm err}(\sigma^2_{{\rm NXS},\Delta f})=\sqrt{\left(\frac{{\rm PN_{lev}}\times \Delta f}{\sqrt{N'}}\right)^2+\frac{2{\rm PN_{lev}}\times \Delta f \sigma^2_{{\rm NXS,}\Delta f}}{N' }} \nonumber \label{err_sigma2}
\end{equation}
\begin{equation}
{\rm err(\sigma_{\rm rms,}}_{\Delta f})=\sqrt{\sigma^2_{{\rm NXS},\Delta f}+ {\rm err}(\sigma^2_{{\rm NXS},\Delta f})}-\sigma_{\rm {rms,}\Delta f}
\end{equation}
Where $N'$ is the number
of periodogram points in $\Delta f$ and
$\sigma^2_{{\rm NXS},\Delta f}$ is the normalised PDS integrated over
$\Delta f$.

\begin{figure}
\psfig{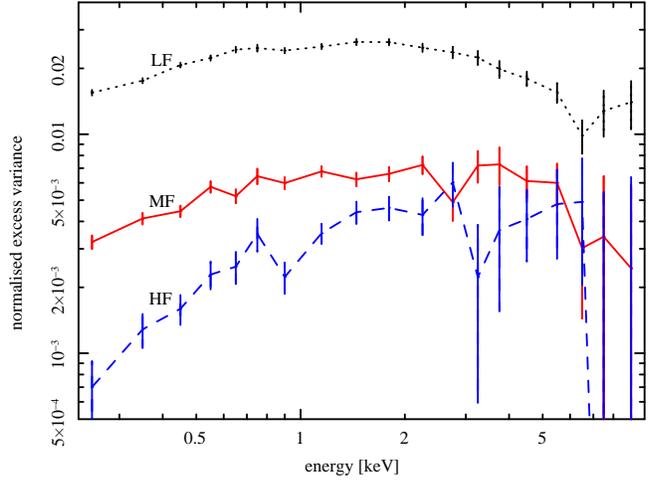}
 \caption{$\sigma^2_{\rm NXS}$ spectra of \mkn\ calculated using the combined data of orbits 999--1003. The low frequency variability ($10^{-5}-10^{-4}$ Hz), in shown in dotted lines,  medium frequency ($10^{-4}-10^{-3.5}$ Hz) in solid lines and high frequency ($10^{-3.5}-10^{-3}$ Hz) in dashed lines.  The error bars account only for Poisson noise in the light curves and were calculated with Eq. \ref{err_sigma2}.}
\label{NXS_999_1003}
\end{figure}

\begin{figure}
\psfig{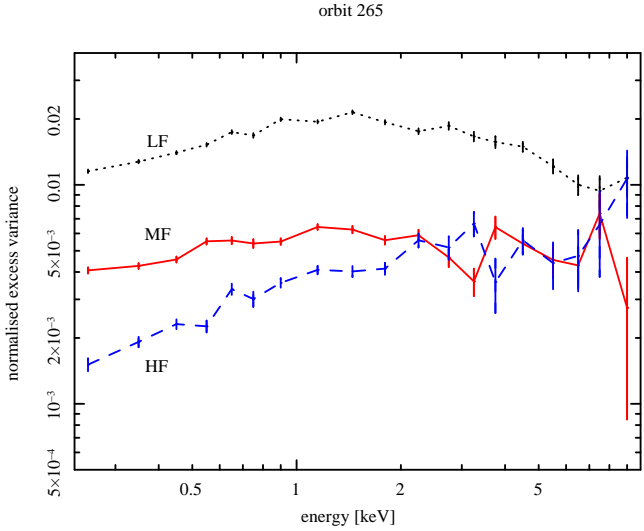}
 \caption{Same as Fig. \ref{NXS_999_1003} but using data from orbit 265 only.}
\label{NXS_265}
\end{figure}

\begin{figure}
\psfig{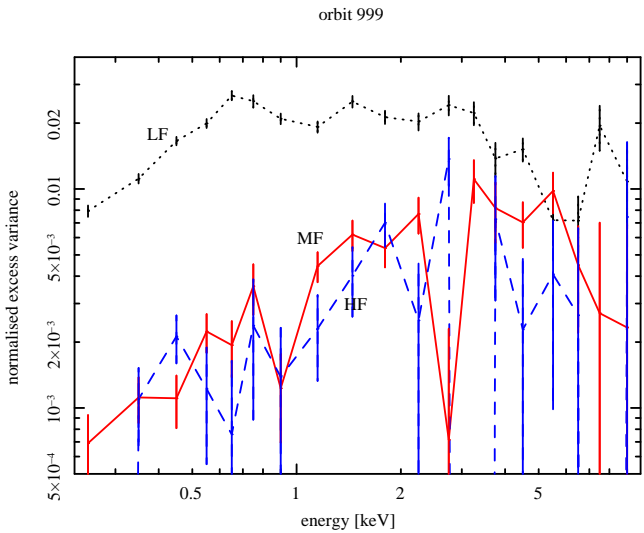}
\caption{Same as Fig. \ref{NXS_999_1003} but using data from orbit 999 only. This data has the lowest count rate of all orbits and displays a much harder $\sigma^2_{\rm NXS}$ medium frequency spectrum. }
\label{NXS_999}
\end{figure}

\section{Normalised excess variance spectra}
\label{nxs}

The $\sigma^2_{\rm NXS}$ spectra of the combined 999--1003 orbit data
are shown in Fig. \ref{NXS_999_1003} and the corresponding spectra of
orbit 265 are shown in Fig. \ref{NXS_265}. In both observations,
separated by 4 years, the source shows similar energy and time-scale
dependencies. The low frequency component has the largest amplitude as
expected, given the $1/f$ shape of the low-frequency PDS and the
logarithmically broader frequency band used for LF. This low-frequency
component, plotted as a dotted line, shows a smooth dip below $\sim 1$
keV and above $\sim 5$ keV, similar to the total fractional
variability spectra of this and other NLS1s \citep{Vaughan766}. At
high frequencies, however, the shape of the $\sigma^2_{\rm NXS}$ is
different, as shown by the dashed line in Figs. \ref{NXS_999_1003} and
\ref{NXS_265}. The HF normalised excess variance shows a stronger dip
at low energies while the high energy dip is reduced or disappears
completely, within the observational uncertainties. Note that, for
each case, the LF and HF \nxs\ spectra have been normalised by the
same time-averaged energy spectrum (i.e. total count rate per energy bin), so
the difference in shape of the LF and HF spectra reveal a time-scale
dependence of the variable spectral component only.

In the two-component interpretation of the main spectral variability,
the energy spectra contains two fixed-shape components which vary in
amplitude. If the two components vary independently, the
variance spectrum can be decomposed as
\begin{equation}
\sigma^2_{{\rm NXS,\Delta} f}(E)= \frac{\sigma ^2_{{\rm rms,\Delta} f,1}(E)+\sigma_{{\rm rms,\Delta} f,2}^2(E)}{(\bar x _{1}(E)+\bar x_{2}(E))^2} 
\end{equation}
where $\sigma_{\rm rms,\Delta f}(E)$ terms represent the rms spectra in a given frequency range $\Delta f$, and $\bar x (E)$ terms are the time-averaged
energy spectra of components 1 and 2.  To produce the curved shape of
the variance spectra (in Figs. \ref{NXS_999_1003}, \ref{NXS_265} and
\ref{NXS_999}), a constant, or less variable, component (1) must
dominate the energy spectrum at low and high energies, therefore
reducing $\sigma^2_{\rm NXS,\Delta f}(E)$ at those energies, while the more
variable component (2) usually assumed to be a power law, might have a
constant $\sigma _{\rm rms,\Delta f}(E)/\bar x _{1}(E)$ throughout the energy
range.  The observed hardening of the $\sigma^2_{\rm NXS,\Delta f}(E)$
spectra towards higher frequencies implies that one or both of the
spectral components, $\sigma ^2_{\rm rms,\Delta f}(E)$ get flatter at higher
frequencies. This can mean that either the variable power law gets
harder at higher frequencies, or that the hard part of `constant'
component appears constant on long time-scales but contributes
significantly to the variance on shorter time-scales. To establish the
energy-spectral shape of the variable components, we will fit the
un-normalised \rms\ spectra in the following section.

The \nxs\ spectra of all individual orbits are qualitatively similar
except for the lowest flux observation, during orbit 999, shown in
Fig. \ref{NXS_999}. In this case, the MF \nxs\ spectrum is notably
different to the LF spectrum. We will examine the variable components
of this observation in detail in Sec. \ref{constant}.  As the data
from orbit 999 appear to behave differently to all other observations,
we will do the spectral fitting analysis to the combined 999--1003 and
1000--1003 data sets separately.

\section{Fourier-resolved \rms\ spectra}
\label{rms}

\begin{figure}
\psfig{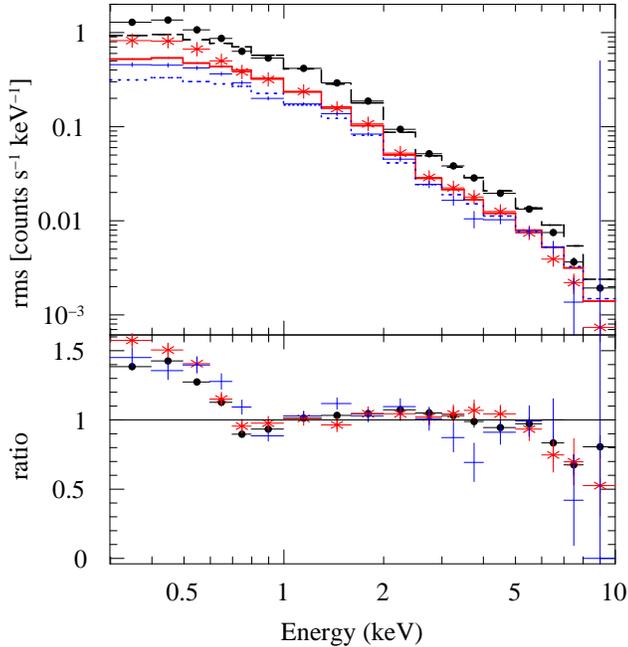}
 \caption{FR spectra of the combined 1000--1003 orbit data, LF in dashed line (black), MF in solid line (red), HF in dotted line (blue). The spectra have been fitted with a power law model at energies above 0.8 keV, and are shown extrapolated down to 0.3 keV. The power law slopes obtained are 1.96 +/-  0.01 for LF, 1.95 +/- 0.02 for MF and 1.76 +/- 0.03 for HF, ($\chi^2/$dof = 4.1 for 33 degrees of freedom). The three temporal-frequency spectra show an apparent `soft excess'.}
\label{rms_999_1003}
\end{figure}

Given the low resolution of the \rms\ spectra we will restrict the
analysis to fitting simple phenomenological spectral components to
obtain the broad shape of the variability components. We start by
fitting the three Fourier frequency \rms spectra of the combined
orbits 999--1003 with a power law, affected only by Galactic
absorption ($n_{\rm H} =1.7 \times 10 ^{20}$ 1/cm$^2$), in the 0.8--10
keV band, giving values for the power law slope of 1.93$\pm$ 0.01 for
LF, 1.85 $\pm$ 0.03 for MF and 1.66 $\pm$0.04 for HF. The extrapolation
of these fits to lower energies shows a `soft excess' over the power
law in the three Fourier-frequency spectra. A similar effect is seen
in the \rms\ spectrum of the combined orbit 1000--1003, where the high
energy power law slopes are 1.96 $\pm$ 0.01 for LF, 1.94 $\pm$ 0.02
for MF and 1.75 $\pm$ 0.04 for HF,($\chi^2/$dof = 3.8 for 33 degrees of
freedom), as shown in Fig. \ref{rms_999_1003}. A similar shape was
found in the FR spectrum analysis of MCG--6-30-15, presented by
\citet{Papadakis_frspec}.

\begin{figure}
\psfig{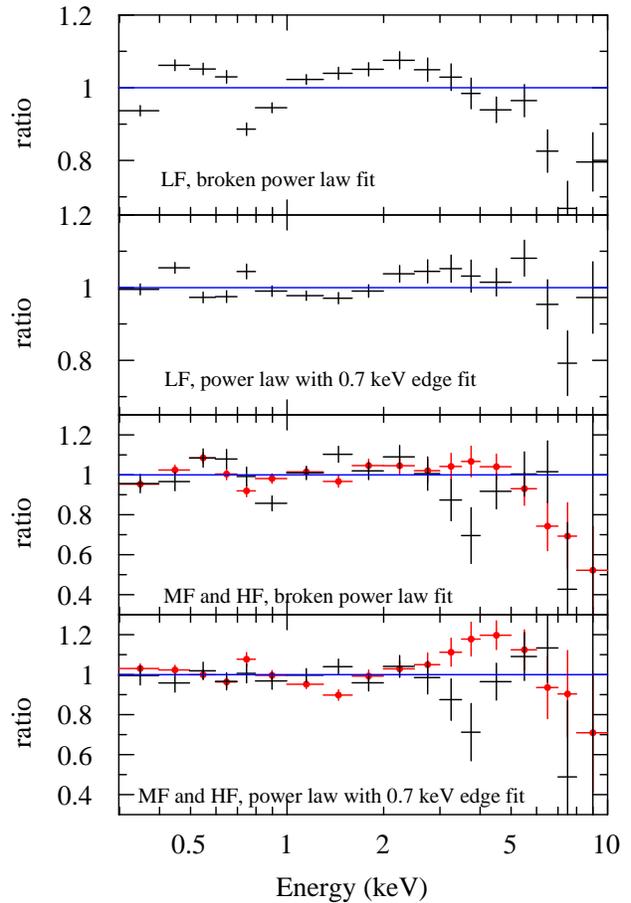}
\caption{Ratio of the LF \rms\ spectrum of orbits 1000--1003, to a broken power law fit (top panel) and to a power law affected by galactic absorption and an absorption edge at $\sim$ 0.7 keV (second panel from top). The third panel shows the residuals of the broken power law fit to the MF (filled dots) and HF (crosses) \rms\ spectra of the same data and the bottom panel shows the MF and HF \rms\ spectra residuals to the warm-absorbed power law model. The power law slopes flatten with increasing frequency: for the warm-absorbed power la model the slopes are $2.14\pm 0.01$ for the LF, $2.19 \pm 0.01$ for the MF and $1.95 \pm 0.02$ for the HF. All fit parameters are given in Tables~\ref{broken_po_table} and \ref{zedge_po_table}. }
\label{edge_po_fit}
\end{figure}

 The soft excess in the \rms\ spectra can be modelled as a broken
power law or as a single power law with an absorption feature around
0.7 keV. We fitted the three frequency ranges with a broken power law
affected by galactic absorption. The resulting fit parameters are
listed in Table~\ref{broken_po_table}. The low and high energy slopes
get flatter with increasing frequency. The low energy slope changes by
approximately 0.15--0.3 and the high energy slope changes by $\sim
0.2$, unlike the case of MCG--6-30-15, where the soft power law showed
no significant frequency dependence \citep{Papadakis_frspec}. The
broken power law model generally does not produce a good fit to the
\rms\ spectra of \mkn\ and the residuals show systematic structure as
shown in the top panel of Fig.~\ref{edge_po_fit}.

\begin{table}
\begin{tabular}{lllll}
&$\Gamma_L$ &$\Gamma_H$&break E& $\chi^2/$dof\\
\hline
999--1003\\
LF&  2.74  $\pm$0.03  & 1.93$\pm$ 0.01& 0.81    $\pm$ 0.02 &13.5  \\
MF&2.69   $\pm$  0.07&1.86 $\pm$  0.03&0.82     $\pm$  0.04&2.90\\
HF&2.39    $\pm$  0.13&1.69    $\pm$  0.04&  0.84     $\pm$  0.1& 2.64\\
\hline
1000--1003\\
LF&  2.62  $\pm$0.04  & 1.95$\pm$ 0.01& 0.77    $\pm$ 0.03 &12.9  \\
MF&2.74   $\pm$  0.06&1.95 $\pm$  0.02&0.79     $\pm$  0.04&3.19\\
HF&2.31    $\pm$  0.10&1.78    $\pm$  0.04&  0.86     $\pm$  0.1& 3.01\\
\hline
265\\
LF&  2.79  $\pm$0.05  & 2.1$\pm$ 0.02& 0.73    $\pm$ 0.03 &6.30  \\
MF&2.83   $\pm$  0.08&2.07 $\pm$  0.03&0.78     $\pm$  0.05&2.28\\
HF&  2.67  $\pm$0.13  & 1.88$\pm$ 0.03& 0.76    $\pm$ 0.07 &1.37  \\
\end{tabular}

\caption{\label{broken_po_table}Fit parameters, with 1$\sigma$ errors, for a broken power law with galactic absorption fit to the 0.3--10 keV FR spectra.}
\end{table}

A better fit to the \rms\ spectra is obtained with a single power law
affected by cold Galactic absorption and a warm absorber. Given the
low resolution of the spectra presented here, we initially simply
modeled the warm absorber as two edges, for O VII and O VIII, using
the Xspec model zedge. The inclusion of the second absorption edge,
however, does not improve the fits significantly, so we dropped one of
the edge components and used a free energy for the remaining edge
instead, to simulate the entire absorption feature. The best fit
parameters to the combined orbits 999--1003, 1000--1003 and orbit 265
are shown in Table \ref{zedge_po_table}. This model produces
systematically lower $\chi^2$ values than the broken power law model,
for the same number of free parameters. The data to model ratio plots
of the 1000-1003 \rms\ spectra fitted with this model are shown in the
second (LF) and fourth (MF and HF) panels of
Fig.~\ref{edge_po_fit}. An absorption edge at 7 keV, improved the fit
in both the broken power law and warm-absorbed power law cases,
however, given the low resolution of the spectra and the large error
bars in the high energy region, these improvements are not
statistically significant.

\begin{table}
\begin{tabular}{lllll}
&edge E &max $\tau$ & Photon Index& $\chi^2/$dof\\
\hline
999--1003\\
LF&  0.80  $\pm$0.01  & 0.71$\pm$ 0.03& 2.17    $\pm$ 0.01 &7.87  \\
MF&0.80   $\pm$  0.01&0.68 $\pm$  0.06&2.11    $\pm$  0.02&3.86\\
HF&0.85    $\pm$  0.02&0.74 $\pm$  0.1&  1.89    $\pm$  0.03& 1.17\\
\hline
1000--1003\\
LF&0.77   $\pm$0.01  & 0.59$\pm$ 0.03& 2.14    $\pm$ 0.01 &3.40 \\
MF&0.76  $\pm$0.01  & 0.62$\pm$ 0.05& 2.19    $\pm$ 0.01 &3.44 \\
HF&0.85  $\pm$0.02  & 0.56$\pm$ 0.07& 1.95    $\pm$ 0.02 &0.97 \\
\hline
265\\
LF&0.75  $\pm$0.01  & 0.55$\pm$ 0.03& 2.28    $\pm$ 0.01 &2.36 \\
MF&0.76  $\pm$0.01  & 0.64$\pm$ 0.06& 2.30    $\pm$ 0.02 &1.36 \\
HF&0.78  $\pm$0.02  & 0.57$\pm$ 0.08& 2.09    $\pm$ 0.02 &1.67 \\
\end{tabular}

\caption{\label{zedge_po_table}Fit parameters, with 1 $\sigma$ errors,  for a power law with galactic absorption and one edge with free energy, fit to the 0.3--10 keV FR spectra.}
\end{table}

A third model, composed of a power law plus black body component, both
affected by galactic absorption, was fitted to the \rms\ spectra,
where the black body component was used to represent the ``soft
excess'' feature. This model does not reproduce well the shape of the
soft excess, producing residuals with structure similar to those of
the broken power law model. The resulting $\chi^2$ values are
systematically worse than for the other two models, for all orbits and
frequency ranges, having values of $\chi^2/$dof = 28.4, 14.0 and 7.3
for the LF spectra of orbit 999--1003, 1000--1003 and 265 data sets,
respectively.

The 0.3--10 keV \rms\ spectra are better fitted by a power law
affected by cold and warm absorption, rather than two power law
components joining at $\sim0.7$ keV or a power law plus a black body
component.  In the power law with 0.7 keV absorption edge model fit,
the spectrum hardens toward higher frequencies with a change in power
law slope of $\sim 0.2$ between LF and HF, while the edge energy and
optical depth remain consistent. The fact that the frequency
dependence of both the low and high energy power law slopes can be
explained by the flattening of a single power law provides further
support to the absorbed power law interpretation.

Table \ref{orbitfits} contains the power law with 0.7 keV absorption edge fit to the LF, MF and HF spectra of each individual orbit. The LF and MF power law slopes of each orbit are equal, while the HF slope is flatter by $\sim 0.2-0.3$ . The low flux observation, orbit 999, is an exception to this rule, in this case the power law slope changes by 0.6 between LF and MF while the HF is similar to the MF spectrum. This observation also produces the largest optical depths for the 0.7 keV absorption edge. 

\begin{table}
\begin{tabular}{l|c|c|c|l}
Orbit &Photon index&edge E&$\tau_{\rm max}$& $\chi^2$/dof\\
\hline
\multicolumn{2}{l}{Low Frequency} \\
\hline
82&2.33 $\pm$ 0.04& 0.83 $\pm$ 0.02&0.90 $\pm$ 0.13& 1.84\\
265 & 2.28 $\pm$0.01& 0.75  $\pm$0.01&  0.55 $\pm$0.03& 2.36\\
999& 2.21 $\pm$ 0.03& 0.85  $\pm$0.01& 1.37  $\pm$0.11& 7.66\\
1000&2.14$\pm$0.01& 0.77  $\pm$0.01& 0.68  $\pm$0.03& 4.03\\
1001&2.24$\pm$0.02& 0.77  $\pm$0.02& 0.41  $\pm$0.06& 1.28\\
1002&2.15$\pm$0.01& 0.78  $\pm$0.01& 0.59  $\pm$0.05& 3.04\\
1003&2.17$\pm$0.02& 0.75  $\pm$0.01&  0.52 $\pm$0.06& 1.59\\
\hline
\multicolumn{2}{l}{Medium Frequency} \\
\hline
82 & 2.28 $\pm$0.05& 0.78  $\pm$0.03& 0.65  $\pm$0.16& 0.67\\
265 & 2.30 $\pm$0.02& 0.76  $\pm$0.01& 0.64  $\pm$0.06& 1.36\\
999 & 1.66 $\pm$0.07&  0.87 $\pm$0.03&1.84 $\pm$0.46& 1.90\\
1000 &2.22  $\pm$0.03& 0.75  $\pm$0.02& 0.58  $\pm$0.09& 0.48\\
1001 &2.11  $\pm$0.03& 0.79  $\pm$0.02& 0.58  $\pm$0.10&1.61 \\
1002 &2.16  $\pm$0.03& 0.82  $\pm$0.01& 0.82  $\pm$0.10& 1.77\\
1003 & 2.20 $\pm$0.02& 0.72  $\pm$0.02& 0.57  $\pm$0.07& 1.79\\
\hline
\multicolumn{2}{l}{High Frequency} \\
\hline
82 & 1.80 $\pm$0.11& 0.83  $\pm$0.06& 0.89  $\pm$0.49&0.99 \\
265 & 2.09 $\pm$0.02& 0.78  $\pm$0.02&  0.57 $\pm$.08& 1.67\\
999 & 1.75 $\pm$0.1& 0.84  $\pm$0.03& 2.67  $\pm$0.89& 1.46\\
1000 & 1.83 $\pm$0.04& 0.86  $\pm$0.05& 0.30  $\pm$0.14& 0.84\\
1001 & 1.81 $\pm$0.05& 0.84  $\pm$0.04& 0.57  $\pm$0.18& 1.05\\
1002 &1.97  $\pm$0.04& 0.86  $\pm$0.02& 0.77  $\pm$0.13& 0.74\\
1003 & 2.07 $\pm$0.04& 0.82  $\pm$0.03& 0.62  $\pm$0.13& 0.92\\
\hline
 \end{tabular}
\caption{\label{orbitfits} Fit parameters to the LF, MF and HF \rms\ spectra of the individual orbit-long observations. the model used was a power law with galactic absorption and an additional absorption edge.}    
\end{table}

\section{\rms\ spectra of the lowest flux observation}
\label{constant}

\begin{figure}
\psfig{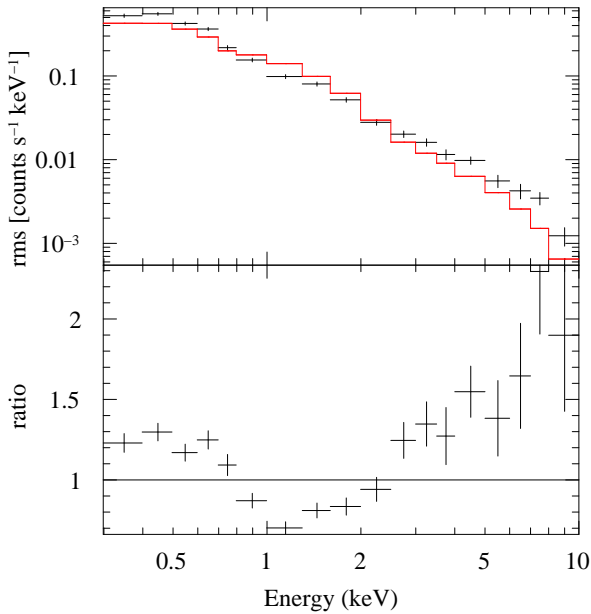}
\caption{Top: LF \rms\ spectrum of the lowest flux observation, orbit 999, in black markers with error bars, the solid line represents the warm-absorbed power law model fit to the combined orbits 1000--1003 LF data and variable normalisation. The bottom panel shows the data to model ratio, showing excess rms variability in the `soft excess' and high energy parts of the spectrum.}
\label{999LF_comp}
\end{figure}

The lowest flux observation, orbit 999, LF \rms\ spectrum produces the
worst fit to the warm-absorbed power law model. Closer inspection of
the 999 LF spectrum reveals excess rms variability around the soft
excess and at the high energy end, compared to the LF spectrum of all
the other orbits. Figure~\ref{999LF_comp} compares the orbit 999 LF
spectrum to the (rescaled) best-fitting model to orbits 1000--1003,
the bottom panel shows the data to model ratio. The orbit 999 LF
spectrum can be fitted by adding a black body component to the
warm-absorbed power law, where the black body acts as a simple
parametrisation of the soft excess. This approach produces a good fit
($\chi^2$/dof=17.92/14=1.28), with parameters kT=0.113$\pm$0.004 keV
, Photon Index =1.55$\pm$0.03 and 0.7 keV edge parameters fixed to the
best-fitting values for orbits 1000--1003. The power law slope is very
flat compared to the LF rms spectra of the other observations, which
have a value of Photon Index $\sim 2.2$ (see Table
\ref{orbitfits}). Alternatively, it is possible that the variable
component in this observation is affected by stronger warm absorption
than the other orbits. To explore this possibility we used the
spectral components identified in these data by \citet{miller,turner}
using higher spectral resolution. In one of their interpretations of
the spectral variability, the most variable components are a power law
of slope $\Gamma=2.38$ plus an ionised reflection component whose
normalisation varies tied to that of the power law, both under a warm
absorber of column density $N_H=2.4\times 10 ^{21} {\rm cm}^{-2}$ and
ionisation parameter $\xi=18.2$. The reflection component is modelled
by the `reflion' model of \citet{ross}, with solar Fe abundance and
$\xi=1600$ erg cm/s. Fitting this model, allowing the normalisations
and power law slope to vary, does not produce a good fit to the LF
\rms\ spectrum of this orbit ($\chi^2/$dof = 7.5, for 15 dof) and
requires a very flat power law of slope 1.36$\pm 0.08$. Allowing the
absorbing column to vary produces a significantly better fit with
$\chi^2/$dof = 4.2, for 14 dof with $N_H=6.5 \pm 0.07 \times 10 ^{21}
{\rm cm}^{-2}$ and power law slope 1.77$\pm0.09$. Fitting this model
to the LF and MF \rms\ spectra simultaneously, with free
normalisations for the power law and reflion components, produces a
reasonably good fit, with $\chi^2/$dof = 2.89, for 30 dof. In this
interpretation, the difference between the LF and MF \rms\ spectra in
this orbit is mainly produced by a decrease in relative normalisation
of the reflected component on short time-scales. The ratio of power
law to reflion normalisation rises from 1.86$\times10^4$ in the LF to
8.37$\times10^4$ in the MF.
                      
The 1--10 keV range of the energy spectrum of orbit 999 is probably
not a single component, as Fig. \ref{NXS_999} shows that the
fractional variability of the LF spectrum in this range changes from
being uniform to dropping sharply at around 3 keV. A natural
explanation for the \rms\ and fractional variability spectra of orbit
999 is that the very strong soft excess and hard power law in the energy
spectrum of this orbit are weakly varying, diluting part of the
fractional variability at low and high energies but still contributing
significantly to the total rms. In this interpretation of the \rms\ 
spectrum, orbit 999 might be essentially equal to all the other
observations with the only difference that the power law component
flux is weak compared to strong-flux, weakly varying additional
components. If this is the case, then the difference between the LF
and MF \rms\ spectra of orbit 999 should be due to the variability
properties of these additional components only, as in all other
observations the LF and MF spectra have approximately the same shape.

We fitted the LF and MF 999 \rms\ spectra with the best-fitting
warm-absorbed power law model of orbits 1000--1003, with all
parameters fixed except for the normalisation, plus a black body
component, and an additional hard power law of Photon Index=1 with an
absorption edge at $\sim$7 keV. These additional components are a simple
parametrisation of the 0-point in the principal component analysis of
\citep{miller}, where the $\sim 7$ keV edge is detected
significantly. We did not use a single absorbed or reflected
component for the additional soft excess and hard power law since
these features show different time-scale dependencies. Both absorption
edges, at 0.7 and 7 keV were applied to all components. The optical
depth of the 7 keV edge and the black body temperature were fitted
jointly to the LF and MF spectra and the normalisations of all
components were allowed to vary independently. Note however that the
\rms\ variability of independent components should add quadratically
and our fitting procedure adds components linearly. This is only a
problem in energy bins where the different components have similar
amplitudes, so we expect to obtain the approximate normalisation of
the different components but not to reproduce the spectral shape
exactly. 

\begin{figure}
\psfig{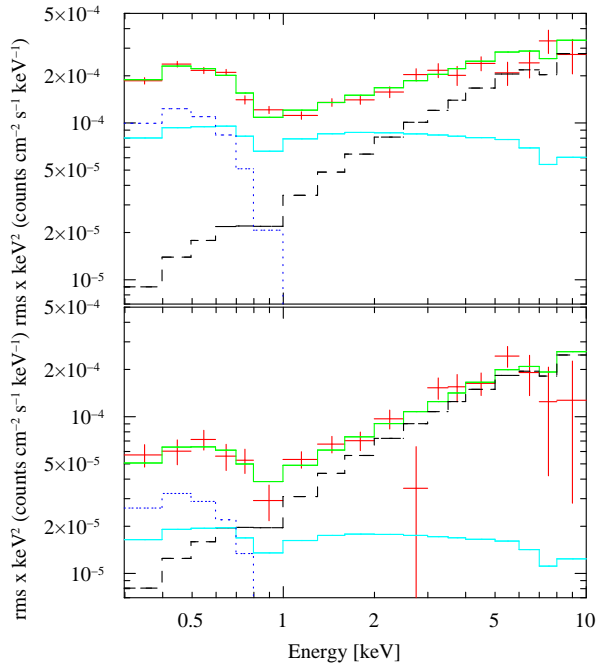}
\caption{LF (top) and MF (bottom) unfolded \rms\ spectra of the lowest flux observation, orbit 999, in red markers with error bars, fitted with the best fitting absorbed power law model to orbits 1000--1003 in light blue solid line, plus a simple parametrisation additional soft excess,represented by a black body (blue dotted line) and a hard power law (black dashed line). In this interpretation, the difference in shape of the LF and MF 999 \rms\ spectra is explained by the relative weakening of the soft power law (as seen in the other orbits) together with a weakening of the soft excess, while the hard power law component remains equally variable in both frequency ranges.}
\label{999LFMF}
\end{figure}

The top panel of Fig.~\ref{999LFMF} shows the 999 LF \rms\ spectrum,
fitted with a black body component and hard power law in addition to
the warm-absorbed power law model fitted to the other orbits, showing
that it can describe the LF spectrum well.  The bottom panel in this
figure shows the same model components fit to the 999 MF spectrum. The
combined fit to the LF and MF spectra produces a
$\chi^2$/dof=38.8/27=1.44. Fitting the same components to the average
spectrum of this orbit, allowing only the normalisations to vary, we
can estimate the fractional variability of the black body component as
rms/$\bar x=7\pm1$\% at low frequencies and $1.9\pm1.0$\% at medium
frequencies, the fractional variability of the hard power law however
remains constant, $7.8\pm1.4$\% fot LF and $7.9 \pm 1.7$\% at MF,
indicating that the additional soft excess and hard power law
components behave differently on different time-scales and cannot be
part of a single, rigid component.  The fact that this additional soft
excess is more variable on long time-scales might be a product of the
soft excess responding to the large amplitude continuum
variations.

The orbit 999 HF \rms\ spectrum is similar to the
corresponding MF spectrum but with much poorer signal-to-noise, given
the low variability power in the high frequency band and the low count
rate in this orbit, so we did not attempt a complex fit to these data.

\section{discussion}

\label{discussion}

We have calculated the normalised excess variance spectra, \nxs , of
Mkn 766, in three different time-scale ranges. The low-frequency
normalised variance spectra show the characteristic shape seen in
other NLS1s, where the fractional variability peaks at energies of
1--3 keV. The drop in fractional variability below 1 keV and above 3
keV can indicate the presence of less variable components, that
dominate the spectrum in those energy ranges, or alternatively,
varying amounts of warm absorption, enhancing the variability at
energies where the opacity is largest. For the first time in AGN X-ray
data analysis, we calculate the normalised excess variance spectrum
for different frequency ranges, finding that this characteristic
peaked shape is only produced by long time-scale fluctuations. We show
that on shorter time-scales (around and above the break frequency in
the PDS) the fractional variability drops strongly at low energies,
while at high energies the drop becomes less pronounced or disappears
completely. We then used the un-normalised \rms\ spectra,
i.e. Fourier-resolved spectra, to study which spectral components
produce this time-scale dependence in the \nxs\ spectra.

The most variable component in the energy spectrum of \mkn\ appears in
the \rms\ spectra as a power law affected by cold (Galactic) and warm
absorption, which give it the appearance of a power law plus a soft
excess. At frequencies around the break frequency in the PDS, the
\rms\ spectrum hardens by 0.2 in power law slope, this behaviour is
observed in two independent observations of the source, in orbit 265
and the combined 1000--1003, separated by 4 years. The fact that the
hardening of both the `apparent soft excess' and the power law in the
\rms\ spectra are consistent with a change in slope of a single
absorbed power law strengthens the interpretation that both features
correspond to a single component.  Therefore, the soft excess seen in
the average energy spectrum can be composed of both, (constant) warm
absorption on a power law component and an additional, less variable
soft component that produces the drop in the \nxs\ spectrum. 

This interpretation is in agreement with the scattering scenario in
the time-resolved spectral analyses of \citet{miller,turner}, where
the spectral variability is mainly explained by a variable power law
continuum and approximately constant scattered
component. \citet{miller} have used principal component analysis to
show that the spectral component producing most of the flux
variability (PC1) is not a simple power law but it contains an
apparent soft excess, caused by the presence of a low column of
absorbing gas ($N_H\sim 3 \times 10^{21} {\rm cm} ^{-2}, \log(\xi)\sim
1$, with $\xi$ in units of erg cm/s), and an Fe line and edge.  This
absorber, which roughly corresponds to the $N_H(3)$ component in the
scattering scenario in \cite{turner}, is consistent with the warm
absorber on the power law in our fits to the \rms\ spectra.  Below, we
speculate on the origin of the observed flux variability and on the
time-scale dependence of the \rms\ spectra.

\subsection{Variable absorption or intrinsic continuum variability}

The spectral variability in these data has been studied using
principal component analysis \citep{miller} and time-resolved spectral
fitting \citep{turner}. These analyses show that the spectral
variability is either produced by a warm absorber covering a variable
fraction of the continuum power law emission, affecting mainly the
flux in the 1--3 keV range, or by a warm-absorbed power law varying in
normalisation in addition to a less variable component containing a
soft excess and a hard power law. Both scenarios produce larger
fractional variability in the 1--3 keV band which can reproduce the
normalised excess variance spectra at low frequencies.

We note that the parameters of the absorbers that produce the
variability in the partial covering model, found by \citet{turner},
can produce a maximum to minimum flux ratio of $\sim 1.5$ in the 5--10
keV band, from uncovered to totally covered continuum. The maximum to
minimum flux ratio in this band in the data (using 25 ks bins) is 2.7
within orbits 999-1003 and 3.5 when all the data are included. This
indicates that either the absorption is much stronger than what is
currently resolved (as, for example, Compton-thick `bricks' in the line
of sight producing additional variability in the entire energy range)
or that the main source of variability in the 5--10 keV band is
largely intrinsic to the continuum flux. The variations on time-scales
of $\sim 25 $ ks are well correlated between different energy bands,
although they are larger at medium energies. In the case where the
5--10 keV variability is mainly driven by intrinsic continuum
variations, this correspondence would imply that either most of the
variability in the entire energy range is produced by the same
continuum variations, or that the strength of the absorption is well
correlated to these variations. The latter case is possible if the
changes in the absorption strength are produced by changes in the
ionisation parameter, instead of covering fraction. The ionisation
state can easily depend on incident flux and, as an increase in flux
can lead to higher ionisation and therefore lower opacity and even
higher observed luminosity, this effect can easily produce the desired
effect of enhanced variability in the 1--3 keV range, where the
warm-absorber opacity is highest.

If the flux variability is intrinsic to the continuum, the larger
fractional variability at medium energies can be produced by constant
or less variable spectral components diluting the fractional
variability at the extremes of the observed energy band. In this
scenario, all the timing properties can be essentially contained in
the flux variability of the continuum, which can be achieved through
fluctuations in the accretion rate, as explained in
Sec.~\ref{prop}. In this case, an approximately constant warm ab sober
covering the varying continuum can produce the \rms\ spectra of the
observed shape. This case would correspond to the `scattering
scenario' for spectral variability described in e.g. \citet{turner},
where the \emph{fractional} variability spectrum requires the presence
of a less variable scattered component.

\subsection{Origin of the time-scale dependence of the \rms\ spectra}

The \rms\ spectra steepen by about 0.2 in power law slope between the
MF and HF regimes. This steepening can be intrinsic to the varying
power law continuum, if for example softer X-rays are emitted by a
larger region which therefore quenches the high-frequency variability,
making the HF spectrum look harder. Notice that this scenario does not
produce nor imply spectral pivoting. Alternatively, approximately
constant absorption can change the slope of the \rms\ spectra if the
region producing high frequency variability is more strongly absorbed
and therefore shows a harder spectrum. In either case, the primary
emitting region needs to be extended and produce different time-scales
of variability in different regions, as for example, variability
produced by the accretion flow on radially-dependent time-scales. The
first case would also require that the emitted spectrum harden toward
the centre of the system, while the second requires larger columns of
absorbing material and/or higher covering fractions obscuring the
innermost regions, where the HF variability would be produced. The HF
spectra covers timescales between 1000 and 3200s, for a black hole
mass of $3.5\times 10^6 M_\odot$ \citep{botte}, 3000s corresponds to a
light crossing time of 200 gravitational radii ($R_g=GM/c^2$) and to
the orbital period at $\sim 10 R_g$. If intrinsic to the primary
continuum, the HF variability is probably produced within a region of
this size, so if constant absorption is producing the change in \rms\
spectral slope, the additional obscuring material has to cover only a
region of a few tens of $R_g$.

In the case where the variability is mainly driven by changes in the
intrinsic continuum flux, a steepening of the \rms\ spectra could
still be produced by some amount of variable absorption. Variable
absorption increases the rms variability mostly in the energy bands
where the absorption is strongest, producing \rms\ spectra with a
shape similar to the \emph{reciprocal} of the opacity as a function of
energy. As mentioned above, the absorbing material parameters found
for this source do not affect strongly the high energy band (5--10
keV), therefore, this process cannot enhance the high energy
variability without enhancing the variability at lower energies much
more. As a consequence, variable absorption can only make the \rms\
spectra softer. To fit the observed behaviour, this process would have
to enhance the low energy variability on long time-scales, covered by
the LF and MF spectra and not the short time-scale variability,
therefore producing comparatively harder HF \rms\ spectra. Notice
however, that the time-scale where variable absorption should start
becoming important needs to coincide with the time-scale at which the
power spectrum of the continuum variability breaks, to produce the
change in slope of the \rms\ spectra at the correct time-scale.

Finally, it is also possible that the hardening of the \rms\ spectra
between the LF and HF regimes is produced by a separate variable
spectral component, with a hard energy spectrum and strong
high-frequency variability. In this interpretation, the power law
spectral component can have the same slope at all variability
time-scales and the time-scale dependence we observe in the \rms\
spectra would be produced by the stronger variability of a hard
component at frequencies of $3 \times 10^{-4}-10^{-3}$ Hz. The
spectrum of this additional component would then correspond to the
spectrum of the possible QPO detected in the PSD of \mkn\ at
$4.6\times 10 ^{-4}$ Hz \citep{markowitz}. The present data are
insufficient to distinguish between an intrinsic hardening of the
power law and the appearance of a separate variability component. We
note however, that it is unlikely that this possible hard component
corresponds to the reflection component. Light travel times to the
reflector will smooth short time-scale variability which is
inconsistent with this hard component displaying \emph{more
  high-frequency power} than the incident continuum. 

\subsection{Propagating fluctuations}
\label{prop}
One interpretation of the change in slope of the \rms\ spectra with frequency is that
fluctuations on different time-scales originate in different regions,
each one emitting a power law energy spectrum with its characteristic
slope. Given that, for each observation, the absorption parameters of
the \rms\ spectra at different frequencies are similar, it is possible
that all these emitting regions are covered by the same (approximately
constant) absorber. This configuration is expected if the power law
component is emitted by a corona close to the central black hole, with
absorbing material, such as a wind, in the line of sight to the
centre.

The behaviour of the \rms\ spectrum can be understood in the context
of propagating-fluctuation models, where the emitted spectrum hardens
towards the centre. Propagating-fluctuation models relate the
variability to accretion rate fluctuations, that are introduced at
different radii in the accretion flow with a radially-dependent
characteristic time-scale, e.g. proportional to the local viscous
time-scale. This scenario was introduced by \citet{lyubarskii} to
explain the wide range of variability time-scales present in the X-ray
light curves of accretion powered systems, and explains several other
variability properties \citep{arevalo}. \citet{kotov} have shown that
if the emitted energy spectrum hardens inward, then this model
produces more high-frequency power at higher energies and also time
lags, where harder bands lag softer ones. 

\citet{churazov} note that a standard geometrically thin accretion
disc cannot produce and propagate short time-scale accretion rate
fluctuations due to its long characteristic time-scales. A
geometrically thick accretion flow, however, can easily maintain and
propagate these rapid fluctuations. In their proposed configuration,
the thermally-emitting geometrically thin disc is sandwiched by a
thick accretion flow, which acts as an accreting corona, responsible
for the Comptonised X-ray emission seen as the power law spectral
component in galactic black hole candidates and AGN. Therefore, a
stable disc emission and highly variable power law emission can
coexist, where the variability arises from accretion rate fluctuations
in the corona. Note that in a standard accretion flow the drift
velocity scales with the scale height squared, $(H/R)^2$, if the
corona is 10 times thicker than the geometrically thin disc and its
surface density is only 1 \% that of the thin disc, the accretion rate
of both flows are comparable. This simple argument shows that even a
tenuous corona can have a significant impact on the total accretion
rate and so it is not unreasonable to think that it can produce most
of the observed X-ray flux variability in AGN.

This interpretation, where fluctuations propagate inwards in the
accretion flow and the spectrum hardens in the same direction,
produces hard lags \citep{kotov, arevalo}, as observed in these data
\citep{markowitz}. In this scenario, increasingly shorter variability
time-scales are produced closer to the centre, and therefore, the
corresponding high-frequency \rms\ spectrum shows a harder energy
spectrum, characteristic of its region of origin.

\label{lastpage}
\end{document}